\documentclass[english]{article}
\usepackage[T1]{fontenc}
\usepackage[latin9]{inputenc}
\usepackage[a4paper]{geometry}
\usepackage{amsmath}
\usepackage{amssymb}
\usepackage{setspace}
\PassOptionsToPackage{normalem}{ulem}
\usepackage{ulem}
\makeatletter

\DeclareRobustCommand{\greektext}{%
  \fontencoding{LGR}\selectfont\def\encodingdefault{LGR}}
\DeclareRobustCommand{\textgreek}[1]{\leavevmode{\greektext #1}}
\DeclareFontEncoding{LGR}{}{}
\DeclareTextSymbol{\~}{LGR}{126}
\newcommand{\lyxmathsym}[1]{\ifmmode\begingroup\def\b@ld{bold}
  \text{\ifx\math@version\b@ld\bfseries\fi#1}\endgroup\else#1\fi}

\newcommand{\lyxaddress}[1]{
\par {\raggedright #1
\vspace{1.4em}
\noindent\par}
}

\makeatother

\usepackage{babel}
\begin{document}

\title{Quaternion-Octonion Unitary Symmetries and Analogous Casimir Operators }

\author{Pushpa\textsuperscript{(1)}, P. S. Bisht\textsuperscript{(1)},
Tianjun Li$^{(2)}$ and O. P. S. Negi$^{(1.2)}$%
\thanks{Address from Feb. 21- April 20, 2012:- Institute of Theoretical Physics,
Chinese Academy of Sciences, Zhong Guan Cun East Street 55, P. O.
Box 2735, Beijing 100190, P. R. China.%
} }

\maketitle
\begin{center}

\par\end{center}

\lyxaddress{\begin{center}
\textsuperscript{(1) }Department of Physics\\
 Kumaun University\\
S. S. J. Campus\\
 Almora - 263601 (U.K.), India.
\par\end{center}}

\lyxaddress{\begin{center}
\textsuperscript{(2)} Institute of Theoretical Physics\\
Chinese Academy of Sciences\\
 Zhong Guan Cun East Street 55\\
 P. O. Box 2735\\
Beijing 100190, P. R. China.
\par\end{center}}

\lyxaddress{\begin{center}
Email-pushpakalauni60@yahoo.co.in \\
ps\_bisht 123@rediffmail.com\\
tli@itp.ac.cn\\
ops\_negi@yahoo.co.in
\par\end{center}}
\begin{abstract}
An attempt has been made to investigate the global $SU(2)$ and $SU(3)$
unitary flavor symmetries systematically in terms of quaternion and
octonion respectively. It is shown that these symmetries are suitably
handled with quaternions and octonions in order to obtain their generators,
commutation rules and symmetry properties. Accordingly, Casimir operators
for $SU(2)$ and $SU(3)$  flavor symmetries are also constructed
for the proper testing of these symmetries in terms of quaternions
and octonions.



\newpage{}
\end{abstract}

\section{Introduction}

Division algebras, specifically quaternions and octonions, Jordan
and related algebras, are described \cite{key-1} in a conceited manner
in unified theories of basic interactions. Quaternions were discovered
by Hamilton \cite{key-2} in 1843 as an illustration of group structure
and also applied to mechanics in three-dimensional space. Quaternions
have the same properties as complex numbers with the difference that
the commutative law is not valid in their case. It gave the importance
to quaternions in terms of their possibility to understand the fundamental
laws of physics. Rather, the octonions \cite{key-3,key-4} form the
widest normed algebra after the algebra of real numbers, complex numbers
and quaternions. The octonions are also known as Cayley Graves numbers
and also have an algebraic structure defined on the 8-dimensional
real vector space in such a way that two octonions can be added, multiplied
and divided with the fact that multiplication is neither commutative
nor associative. Nevertheless, other expected properties like distributivity
and alternativity hold well to the case of octonions. Indeed, Pais
\cite{key-5} pointed out a striking similarity between the algebra
of interactions and the split octonion algebra. Furthermore, some
attention has been given to octonions \cite{key-6} in theoretical
physics in order to extend the 3+1 space-time to eight dimensional
space-time as the consequence to accommodate the ever increasing quantum
numbers and internal symmetries related to elementary particles and
gauge fields. However, Günaydin and Gürsey \cite{key-7} has also
developed the quark model and color gauge theory in terms of split
octonions. Above all, a Casimir operator \cite{key-8,key-9}, named
after Hendrik Casimir, was described as an important tool in the study
\cite{key-10,key-11,key-12,key-13,key-14,key-15,key-16} of associative
and alternative algebras. There is an infinite family of Casimir operators
whose members are expressible in terms of a number of primitive Casimirs
equal to the rank of the underlying group. Moreover, an algebra of
colors suitably handled with octonions has been discussed \cite{key- 17}
for two Casimir operators and the two generators of the Cartan sub
algebra of the automorphism group $SU(3)$ of the Hilbert space. Further,
we have developed \cite{key-18} the quaternionic formulation of Yang-
Mill's field equations and octonionic reformulation of quantum chromo
dynamics (QCD) where the resemblance between quaternions and $SU(2)$
and that of octonions and $SU(3)$ gauge symmetry has been discussed.
The color group $SU(3)_{C}$ is embedded with in the octonionic structure
of the exceptional groups while the $SU(3)$ flavor group has been
discussed in terms of triality property of the octonion algebra. Therefore,
we have also derived the relations for different components of isospin
with quark states. On the other hand , we \cite{key-19} have analyzed
this formalism to the case of $SU(3)$ flavor group in terms of its
splitting to various pairs of $SU(2)$ isospin handled with quaternions.
Likewise, the various commutation relations among the generators of
$SU(3)$ group and its shift operators have  been derived and verified
in terms of different isospin multiplets i.e. $I$, $U$ and $V$-
spins. Keeping in view, the utilities of quaternions and octonions
in internal symmetry groups, in this paper, we have made an attempt
to investigate the global $SU(2)$ and $SU(3)$ unitary flavor symmetries
respectively in terms of quaternions and Octonions. Therefore, we
have developed $SU(2)$ flavor symmetry from quaternions to relate
isotopic spin and also $SU(3)$ symmetry (the so called eight fold
way) from octonions to study the three flavors of quarks and anti-quarks.
It is shown that these symmetries are suitably handled with quaternions
and octonions in order to obtain their generators, commutation rules
and symmetry properties. Consequently, the analogous Casimir operators
for $SU(2)$ and $SU(3)$ flour symmetry groups are analyzed and suitably
handled respectively with quaternions and octonions. It is also shown
that analogous Casimir operators commute with the corresponding generators
of $SU(2)$ and $SU(3)$ gauge groups.

\section{Quaternion Gauge Theory and Isospin }

Let $\phi(x)$ be a quaternionic field ($Q$ field) and expressed
as 

\begin{align}
\phi= & e_{0}\phi_{0}+e_{j}\phi_{j}\,\,(\forall\, j=1,2,3)\label{eq:1}
\end{align}
where $\phi_{0}$ and $\phi_{j}$ are local Hermitian fields while
the $e_{0}$ and $e_{j}$ are respectively the real and the imaginary
basis of quaternion \textbf{$Q$ }satisfying the following multiplication
property

\begin{eqnarray}
e_{0}^{2} & =e_{0};\,\, & e_{0}e_{j}=-\delta_{jk}e_{0}+\epsilon_{jkl}e_{l}\,\,\,(\forall j,k,l=1,2,3).\label{eq:2}
\end{eqnarray}
 The quaternion basis elements $(e_{0},e_{1},e_{2},e_{3})$ may also
be written \cite{key-20,key-21,key-22,key-23,key-24,key-25} in terms
of $4\times4$ real or $2\times2$ complex matrices. The $2\times2$
correspondence with complex matrices is described \cite{key-21,key-22}
as

\begin{align}
e_{1}\Rightarrow & \left[\begin{array}{cc}
0 & i\\
i & 0
\end{array}\right];\,\, e_{2}\Rightarrow\left[\begin{array}{cc}
0 & 1\\
-1 & 0
\end{array}\right];\,\, e_{3}\Rightarrow\left[\begin{array}{cc}
i & 0\\
0 & -i
\end{array}\right]\label{eq:3}
\end{align}
which are expressed as $2\times2$ Pauli spin Hermitian matrices $\tau_{j}(\forall j=1,2,3)$
connected to spatial unitary group $SU(2)$ as 

\begin{align}
\tau_{1}=-ie_{1}\Rightarrow & \left[\begin{array}{cc}
0 & 1\\
1 & 0
\end{array}\right];\,\,\tau_{2}=-ie_{2}\Rightarrow\left[\begin{array}{cc}
0 & -i\\
i & 0
\end{array}\right];\,\,\tau_{3}=-ie_{3}\Rightarrow\left[\begin{array}{cc}
1 & 0\\
0 & -1
\end{array}\right].\label{eq:4}
\end{align}
However, in quantum Chromodynamics, flavor is a global symmetry where
the unitary transformations are independent of space and time. Rather,
this symmetry is broken in the electroweak theory and flavor changing
processes exist, such as quark decay or neutrino oscillations. $SU(2)$
isospin flavor symmetry at the quark level is denoted by $SU(2)_{f}$.
Therefore, the $SU(2)$ global gauge symmetry may be handled with
quaternion gauge formalism in an enthusiastic manner. So, a quaternion
spinor $\psi$ transforms as 

\begin{eqnarray}
\psi & \longmapsto\psi^{\shortmid}= & U\,\psi\label{eq:5}
\end{eqnarray}
where $U$ is $2\times2$ unitary matrix and satisfies 

\begin{eqnarray}
U^{\dagger}U & =UU^{\dagger}=UU^{-1}=U^{-1}U= & 1.\label{eq:6}
\end{eqnarray}
On the other hand, the quaternion conjugate spinor transforms as 
\begin{eqnarray}
\overline{\psi}\longmapsto\overline{\psi^{\shortmid}} & = & \overline{\psi}U^{-1}.\label{eq:7}
\end{eqnarray}
Therefore, the combination $\psi\overline{\psi}=\overline{\psi}\psi=\psi\overline{\psi^{\shortmid}}=\overline{\psi^{\shortmid}}\psi$
is an invariant quantity. We may thus write any unitary matrix as 

\begin{eqnarray}
U & = & \exp\left(i\,\hat{H}\right)\,\,(i=\sqrt{-1})\label{eq:8}
\end{eqnarray}
where $\hat{H}$ is Hermitian $\hat{H}^{\dagger}=\hat{H}$. Here,
the Hermitian $2\times2$ matrix may be defined in terms of four real
numbers, $\alpha_{1,}\alpha_{2},\,\alpha_{3}$ and $\alpha_{0}$ as 

\begin{eqnarray}
\hat{H} & = & \alpha_{0}\hat{1}+\tau_{j}\alpha_{j}=\alpha_{0}e_{0}-i\,\alpha_{j}e_{j}\,(\forall j=1,2,3)\label{eq:9}
\end{eqnarray}
where $\hat{1}$ is the $2\times2$ unit matrix and thus we may write
the Hermitian matrix $\hat{H}$ as 

\begin{eqnarray}
\hat{H} & = & \left(\begin{array}{cc}
\alpha_{0}+\alpha_{3} & \alpha_{1}-i\alpha_{2}\\
\alpha_{1}+i\alpha_{2} & \alpha_{0}-\alpha_{3}
\end{array}\right).\label{eq:10}
\end{eqnarray}
For global gauge transformations both $\alpha_{0}=\theta\,\mbox{(say)}$
and $\overrightarrow{\alpha}$ are independent of space-time so that
the proton-neutron (or u-d quark) doublet wave function be described
\cite{key-19} as

\begin{align}
\psi= & \left(\begin{array}{c}
u\\
d
\end{array}\right)\label{eq:11}
\end{align}
where $|u>=\left(\begin{array}{c}
1\\
0
\end{array}\right)$ and $|d>=\left(\begin{array}{c}
0\\
1
\end{array}\right)$ and equation (\ref{eq:11}) then transforms as 

\begin{align}
V\left(\begin{array}{c}
u\\
d
\end{array}\right)V^{-1}\equiv & U\,\left(\begin{array}{c}
u\\
d
\end{array}\right)\label{eq:12}
\end{align}
with

\begin{align}
U=\exp\left(\frac{1}{2}\alpha_{j}\tau_{j}\right)\equiv & \exp\left(-\frac{1}{2}\alpha_{j}e_{j}\right)\equiv\exp(i\,\overrightarrow{\alpha}.\overrightarrow{T}).\label{eq:13}
\end{align}
Here, the isospin $\overrightarrow{T}$ is considered as the analog
of the angular momentum and the rotational invariance in internal
isotopic spin space implies that the isospin is conserved. The generators
of isospin {[}i.e the $SU(2)${]} group are defined as $\hat{T_{j}}=\frac{1}{2}\tau_{j}\equiv\frac{1}{2}\, e_{j}(\forall j=1,2,3)$
which satisfy the following well known commutation relation

\begin{align}
\left[\widehat{T}_{j},\widehat{T}_{k}\right]= & i\epsilon_{jkl}\widehat{T}_{l}\,(\forall j,k,l=1,2,3).\label{eq:14}
\end{align}
In $SU(2)_{f}$ ,define the base state $\psi=\left(\begin{array}{c}
u\\
d
\end{array}\right)\equiv\mathbf{2}$, is sometimes called the \textquotedblleft{}fundamental representation\textquotedblright{}.
Accordingly, the conjugate state corresponds to the antiparticle (ant-quark
or anti-nucleon) states which are described by the complex conjugate
(not the Hermitian conjugate) of the $SU(2)_{f}$ representation since
here we require the transformations for column vector $(\begin{array}{c}
\overline{u}\\
\overline{d}
\end{array})$ instead of the row vector $(u,d)$. So, by identifying $\bar{u}\equiv u^{\star}$
and $\bar{d}\equiv d^{\star}$, we may write $SU(2)_{f}$ transformation
law for the anti quark doublet as 

\begin{gather}
\psi^{\star\shortmid}\cong\left(\begin{array}{c}
\bar{u'}\\
\bar{d'}
\end{array}\right)\Longleftrightarrow U^{\star}\psi^{\star}\cong\exp\left(\alpha_{j}e_{j}/2\right)\left(\begin{array}{c}
\bar{u}\\
\bar{d}
\end{array}\right).\label{eq:15}
\end{gather}
It should be noted that the quark doublet $\left(u,d\right)$ and
anti-quark doublet $\left(\bar{u},\bar{d}\right)$ transform differently
under $SU(2)_{f}$ transformations. These two representations are
unitary equivalent, so we can consider unitary matrix $U_{C}$ as

\begin{align}
U_{C}\exp\left(\alpha_{j}e_{j}/2\right)U_{C}^{-1}= & \exp\left(-\alpha_{j}e_{j}/2\right).\label{eq:16}
\end{align}
Here, the unitary matrix $U_{C}$ satisfies the following conditions
i.e.

\begin{align}
U_{C}\left(e_{1}\right)U_{C}^{-1}= & -e_{1};\nonumber \\
U_{C}\left(e_{2}\right)U_{C}^{-1}= & e_{2};\nonumber \\
U_{C}\left(e_{3}\right)U_{C}^{-1}= & -e_{3}\label{eq:17}
\end{align}
and in order to obtain a convenient unitary representation, one can
choose $U_{C}$ as

\begin{align}
U_{C}= & -e_{2}=\left(\begin{array}{cc}
0 & 1\\
-1 & 0
\end{array}\right).\label{eq:18}
\end{align}
This implies that the doublet

\begin{align}
\psi_{C}=U_{C}\left(\begin{array}{c}
\bar{u}\\
\bar{d}
\end{array}\right)= & \left(\begin{array}{c}
\bar{d}\\
-\bar{u}
\end{array}\right)\equiv\overline{2};\label{eq:19}
\end{align}
transforms exactly in the same way as $\psi=\left(\begin{array}{c}
u\\
d
\end{array}\right)\equiv\mathbf{2}$. This result is useful in order to write the familiar table of (Clebsch-Gordan)
angular momentum \cite{key-13} coupling coefficients for combining
quark and anti-quark states together. For this, we include the relative
minus sign between the $\bar{d}$ and $\bar{u}$ components which
has appeared in equation $\left(\ref{eq:19}\right)$. As such,
\begin{itemize}
\item \textbf{For $\mathbf{T=0,}$} $\bar{q}q$ combination $\frac{1}{\sqrt{2}}(\bar{u}u+\bar{d}d)$$\longmapsto\frac{1}{\sqrt{2}}(u^{\star}u+d^{\star}d)$$\cong\psi^{\dagger}\psi$
where $\frac{1}{\sqrt{2}}$ is used as normalization constant. So,
under an $SU(2)_{f}$ transformation,
\end{itemize}
\begin{align}
\psi^{\dagger}\psi\longmapsto\psi^{\shortmid\dagger}\psi^{\shortmid}= & \psi^{\dagger}U^{\dagger}U\psi=\psi^{\dagger}\psi;\label{eq:20}
\end{align}
showing that $\psi^{\dagger}\psi=\frac{1}{\sqrt{2}}(\bar{u}u+\bar{d}d)\equiv|0,0>$
is indeed an $SU(2)_{f}$ invariant for isospin singlet states $T=0$
and $T_{3}=0$
\begin{itemize}
\item \textbf{For $\mathbf{T=1}$}(isospin triplet), we consider the three
quantities $\varphi_{j}$ defined as
\end{itemize}
\begin{align}
\varphi_{j}= & \psi^{\dagger}\left(ie_{j}\right)\psi\,\,(\forall\, j=1,2,3)\label{eq:21}
\end{align}
along with an infinitesimal $SU(2)_{f}$ transformation 

\begin{align}
\psi^{\shortmid}= & \left(\hat{1}_{2}-\alpha_{j}e_{j}/2\right)\psi.\label{eq:22}
\end{align}
So, on using equations $\left(\ref{eq:11}\right)$ and $\left(\ref{eq:20}\right)$,
we get the iso- triplet for $T=1$ as 

\begin{align}
\left(\begin{array}{c}
\varphi_{1}\\
\varphi_{2}\\
\varphi_{3}
\end{array}\right)\rightleftharpoons\frac{1}{\sqrt{2}} & \left(\begin{array}{c}
(\bar{u}d+\bar{du})\\
(-i\bar{ud}+i\bar{du)}\\
\bar{(u}u-\bar{d}d)
\end{array}\right)\label{eq:23}
\end{align}
where $\varphi_{3}=\frac{1}{\sqrt{2}}(\bar{u}u-\bar{d}d)\equiv|1,0>$
describes the state corresponding to quantum number isospin $(T=1;T_{3}=0)$
analogous to neutral pion $\pi^{0}$. Similarly, we may assign the
isospin quantum numbers to the linear combination of $\varphi_{1}$and
$\varphi_{2}$as for isospin $(T=1;T_{3}=-1)$ analogous to charged
pions $\pi^{-}$as

\begin{align}
\frac{1}{\sqrt{2}}\left(\varphi_{1}+i\varphi_{2}\right) & \rightleftharpoons\bar{u}d\equiv|1,-1>\label{eq:24}
\end{align}
and for isospin $(T=1;T_{3}=+1)$ analogous to charged pions $\pi^{+}$
as 
\begin{align}
\frac{1}{\sqrt{2}}\left(\varphi_{1}-i\varphi_{2}\right) & \rightleftharpoons\bar{d}u\equiv|1,+1>.\label{eq:25}
\end{align}

\section{Quaternions and Dirac spinors}

Let us define the free particle quaternion Dirac equation \cite{key-26}
for a particle of mass $m$ as 

\begin{align}
\left(i\gamma^{\mu}\partial_{\mu}-m\right)\psi= & 0.\label{eq:26}
\end{align}
Here, we used the following quaternion valued Weyl representation
of $\gamma$ matrices as it is convenient for ultra relativistic problems
i.e

\begin{align}
\gamma^{0}=\left[\begin{array}{cc}
0 & e_{0}\\
e_{0} & 0
\end{array}\right],\equiv\left[\begin{array}{cc}
0 & \hat{1}_{2}\\
\hat{1}_{2} & 0
\end{array}\right],\,\,\,\,\,\, & \gamma^{j}=\left[\begin{array}{cc}
0 & -ie_{j}\\
ie_{j} & 0
\end{array}\right]\equiv\left[\begin{array}{cc}
0 & \tau^{j}\\
-\tau^{j} & 0
\end{array}\right](\forall j=1,2,3).\label{eq:27}
\end{align}
In Weyl representation, we have $\left(\gamma^{0}\right)^{\dagger}=\gamma^{0}$,
$\left(\gamma^{j}\right)^{\dagger}=-\gamma^{j}$ and $\left\{ \text{\textgreek{g}}^{\mu},\,\text{\textgreek{g}}^{\nu}\right\} =\text{\textgreek{g}}^{\mu}\text{\textgreek{g}}^{\nu}+\gamma^{\nu}\gamma^{\mu}=0\,(\forall\mu,\nu=0,1,2,3)$.
Rather, at sufficiently low energies \cite{key-27} the effect of
weak interactions dies away and the dominant contribution appears
due to the presence of electromagnetic and the strong interactions
both of which are Parity conserving interactions. So, we use the simplest
representation that closes under Parity $\mathbb{\hat{P}}$ for Dirac
spinor field $\psi$ in equation (\ref{eq:26}) as,

\begin{align}
\psi= & \left(\begin{array}{c}
\psi_{L}\\
\psi_{R}
\end{array}\right)\cong(\frac{1}{2},\,0)\,\oplus(0,\,\frac{1}{2})\label{eq:28}
\end{align}
where $(\frac{1}{2},\,0)$ and $(0,\,\frac{1}{2})$ are the simplest
representations of the Lorentz group $SL(2,C)$ so that an object
transforming in the $(\frac{1}{2},\,0)$ representation is called
a left-chiral $\psi_{L}$ Weyl spinor. Similarly, an object transforming
in the $(0,\,\frac{1}{2})$ representation of Lorentz group $SL(2,C)$,
is called a right-chiral Weyl spinor $\psi_{R}$. It is to be noted
that the meaning assigned to the technical word 'chiral' (or so called
chirality means handedness)) is associated with the matrix $\lyxmathsym{\textgreek{g}}^{5}$
which anti-commutes with all Dirac matrices (\ref{eq:27}) i.e.

\begin{align}
\left\{ \lyxmathsym{\textgreek{g}}^{5}\,,\,\lyxmathsym{\textgreek{g}}^{\mu}\right\} = & \text{\textgreek{g}}^{5}\lyxmathsym{\textgreek{g}}^{\mu}+\gamma^{\mu}\gamma^{5}=0\,\,(\forall\mu=0,1,2,3).\label{eq:29}
\end{align}
So, from the anti-commutation relation $\left\{ \text{\textgreek{g}}^{\mu},\,\lyxmathsym{\textgreek{g}}^{\nu}\right\} =\text{\textgreek{g}}^{\mu}\text{\textgreek{g}}^{\nu}+\gamma^{\nu}\gamma^{\mu}=0\,(\forall\mu,\nu=0,1,2,3)$
between the $\lyxmathsym{\textgreek{g}}$matrices (\ref{eq:27}),
it can be easily seen that the matrix

\begin{align}
\gamma^{5}=i\gamma^{0}\gamma^{1}\gamma^{2}\gamma^{3} & =\left(\begin{array}{cc}
-e_{0} & 0\\
0 & e_{0}
\end{array}\right)\equiv\left(\begin{array}{cc}
-\widehat{1}_{2} & 0\\
0 & \hat{1}_{2}
\end{array}\right)\label{eq:30}
\end{align}
satisfi{}es Eq. (\ref{eq:29}) along with the following properties

\begin{align}
\left(\gamma^{5}\right)^{\dagger}=\gamma^{5};\,\,\,\,\,\,\, & \left(\gamma^{5}\right)^{2}=\hat{1}_{2}\equiv1.\label{eq:31}
\end{align}
As such, we may define the \textquotedblleft{}left-handed\textquotedblright{}
and \textquotedblleft{}right-handed\textquotedblright{} projection
operators (which are generally used as the terms \textquotedblleft{}left-chiral\textquotedblright{}
and \textquotedblleft{}right-chiral\textquotedblright{}) as

\begin{align}
\widehat{P}_{L}=\frac{1}{2}\left(1-\gamma^{5}\right);\,\,\,\,\,\,\,\, & \widehat{P}_{R}=\frac{1}{2}\left(1+\gamma^{5}\right);\label{eq:32}
\end{align}
which shows that the operators $\widehat{P}_{L/R}$ project onto the
left/right-chiral Weyl spinor as

\begin{align}
\widehat{P}_{L}\psi=\frac{1}{2}\left(1-\gamma^{5}\right)\psi=\left(\begin{array}{c}
\psi_{L}\\
0
\end{array}\right)\equiv\psi_{L};\,\,\,\,\,\,\,\, & \widehat{P}_{R}\psi=\frac{1}{2}\left(1+\gamma^{5}\right)\psi=\left(\begin{array}{c}
0\\
\psi_{R}
\end{array}\right)\equiv\psi_{R}.\label{eq:33}
\end{align}
Thus, we may write the suitable Dirac Lagrangian for left-right chiral
spinors as,

\begin{align}
\mathcal{L}_{D}= & \psi_{L}^{\dagger}\overline{e_{\mu}}\partial_{\mu}\psi_{L}+\psi_{R}^{\dagger}\overline{e_{\mu}}\partial_{\mu}\psi_{R}-m\left(\psi_{L}^{\dagger}\psi_{R}+\psi_{R}^{\dagger}\psi_{L}\right);\label{eq:34}
\end{align}
where $\mu=0,1,2,3.$ Here, the Euler-Lagrange equations are obtained
on considering $\psi_{L}$ and $\psi_{L}^{\star}$ independent (since
they are complex fields) and similarly for $\psi_{R}$, $\psi_{R}^{\star}$.
Then, performing the variations with respect to $\psi_{L}^{\star}$
and $\psi_{R}^{\star}$, we get that the Dirac equation (\ref{eq:26})
is equivalent to the pair of equations

\begin{align}
\overline{e_{\mu}}\partial_{\mu}\psi_{L}= & m\psi_{R};\nonumber \\
\overline{e_{\mu}}\partial_{\mu}\psi_{R}= & m\psi_{L}.\label{eq:35}
\end{align}
These two pairs of equations decouple for the $m=0$ (i.e for zero
mass) particles like neutrinos. We may also write the Lagrangian in
a compact form for which it is convenient to define the Dirac adjoint
spinor as $\overline{\psi}=\psi^{\dagger}\gamma^{0}\equiv\left(\psi_{R}^{\dagger},\psi_{L}^{\dagger}\right)$
in chiral representation. hence, the compact and consistent form of
the Dirac Lagrangian may be written as,

\begin{align}
L_{D}= & \overline{\psi}\left(i\gamma^{\mu}\partial_{\mu}-m\right)\psi.\label{eq:36}
\end{align}

\section{Flavor $SU(3)_{f}$ and Octonions}

Gell-Mann and Ne'eman \cite{key-28} were the first to propose $SU(3)_{f}$
as the correct generalization of isospin $SU(2)_{f}$ to include third
quantum number the strangeness along with the isospin. The Gell-Mann
$\lambda$ matrices are used for the representations of the infinitesimal
generators of the special unitary group called $SU(3)$. This group
consists eight linearly independent generators $F_{A}=\frac{\lambda_{A}}{2}(\forall\, A=1,2,3,........8)$
which satisfy the following commutation relation as

\begin{align}
\left[F_{A}\,,F_{B}\right]= & if_{ABC}F_{C}\label{eq:37}
\end{align}
where $f_{ABC}$ is the structure constants of $SU(3)$ like $\epsilon_{jkl}$
of $SU(2)$ and is completely antisymmetric. The Gell-Mann matrices
$\lambda_{A}(\forall\, A=1,2,3,........8)$ are defined as

\begin{align}
\lambda_{1}= & \begin{pmatrix}0 & 1 & 0\\
1 & 0 & 0\\
0 & 0 & 0
\end{pmatrix};\qquad\:\lambda_{2}=\,\begin{pmatrix}0 & -i & 0\\
i & 0 & 0\\
0 & 0 & 0
\end{pmatrix};\nonumber \\
\lambda_{3}= & \begin{pmatrix}1 & 0 & 0\\
0 & -1 & 0\\
0 & 0 & 0
\end{pmatrix};\qquad\lambda_{4}=\,\begin{pmatrix}0 & 0 & 1\\
0 & 0 & 0\\
1 & 0 & 0
\end{pmatrix};\nonumber \\
\lambda_{5}= & \begin{pmatrix}0 & 0 & -i\\
0 & 0 & 0\\
i & 0 & 0
\end{pmatrix};\qquad\lambda_{6}=\,\begin{pmatrix}0 & 0 & 0\\
0 & 0 & 1\\
0 & 1 & 0
\end{pmatrix};\nonumber \\
\lambda_{7}= & \begin{pmatrix}0 & 0 & 0\\
0 & 0 & -i\\
0 & i & 0
\end{pmatrix};\qquad\lambda_{8}=\,\frac{1}{\surd3}\begin{pmatrix}1 & 0 & 0\\
0 & 1 & 0\\
1 & 0 & -2
\end{pmatrix}\label{eq:38}
\end{align}
which satisfy the following properties

\begin{align}
\left(\lambda_{A}\right)^{\dagger}= & \lambda_{A};\nonumber \\
Tr(\lambda_{A})=0\,\,\,\,\,\,\, & Tr(\lambda_{A}\lambda_{B})=2\delta_{AB};\nonumber \\
\left[\lambda_{A},\lambda_{B}\right]= & 2if^{ABC}\lambda_{C};\nonumber \\
\left\{ \lambda_{A},\lambda_{B}\right\} = & \frac{4}{3}\delta_{AB}+2\, d_{ABC}\lambda_{C}.\label{eq:39}
\end{align}
Here $d_{ABC}$ is totally symmetric tensor and it is required only
to obtain one of the Casimir operators of $SU(3)$ symmetry group.
Gellmann $\lambda$ matrices obviously act on three component column
vectors as the generalization of the two component isospinors of $SU(2).$
The generators of $SU(3)_{f}$ connect three quarks namely the up
$(u)$, down $(d)$. and strange $(s)$ quarks so that we may consider
unitary $3\times3$ transformations among them as

\begin{align}
\psi^{\shortmid}= & W\psi;\label{eq:40}
\end{align}
where $\psi$ now stands for the three component column vector

\begin{align*}
\psi= & \left(\begin{array}{c}
u\\
d\\
s
\end{array}\right);
\end{align*}
and $W$ is the $3\times3$ unitary matrix of determinant$1$. The
representation provided by this triplet of states is called the fundamental
representation of $SU(3)_{f}$. An infinitesimal $SU(3)$ matrix may
then be defined as 

\begin{align}
W_{infl}= & \hat{1}_{3}+i\chi;\label{eq:41}
\end{align}
where $\hat{1}_{3}$is the unit matrix of order $3$ and $\chi$ is
$3\times3$ matrix and satisfies the properties of octonions \cite{key- 17,key-19}.
The relation between Gell - Mann $\lambda$ matrices and octonion
units are given \cite{key-18} as 

\begin{align}
e_{1}\Rightarrow & i\lambda_{1},\; e_{2}\Rightarrow i\lambda_{2},\; e_{3}\Rightarrow i\lambda_{3}\longmapsto e_{A}\Longleftrightarrow i\lambda_{A};\,\,\,\,\,(\forall\: A=1,2,3);\nonumber \\
e_{4}\Rightarrow & \frac{i}{2}\lambda_{4},\,\,\, e_{5}\Rightarrow\frac{i}{2}\lambda_{5},\Longleftrightarrow e_{A}=\frac{i}{2}\lambda_{A};\,\,\,\,\,(\forall\: A=4,5,);\nonumber \\
e_{6}\Rightarrow- & \frac{i}{2}\lambda_{6},\text{\,\,\,}e_{7}\Rightarrow-\frac{i}{2}\lambda_{7},\Longleftrightarrow e_{A}=-\frac{i}{2}\lambda_{A};\,\,\,\,\,(\forall\: A=6,7,);\nonumber \\
e_{0}\Longleftrightarrow & \frac{\surd3}{2}\lambda_{8};\label{eq:42}
\end{align}
where $e_{A}(A=1,2,.....,7)$ are imaginary octonion units and $e_{0}$
is the real octonion basis element corresponding to unity. Here, a
set of octets $(e_{0},\, e_{1},\, e_{2},\, e_{3},\, e_{4},\, e_{5},\, e_{6},\, e_{7})$
are known as the octonion \cite{key-6} basis elements and satisfy
the following multiplication rules

\begin{align}
e_{0}=1;\,\, e_{0}e_{A}=e_{A}e_{0}= & e_{A};\nonumber \\
e_{A}e_{B}=-\delta_{AB}e_{0}+f_{ABC}e_{C}.\,\,(\forall\, A,\, B,\, C=1,\,2,.....,\,7) & .\label{eq:43}
\end{align}
The structure constants $f_{ABC}$ is completely antisymmetric and
takes the value $\mbox{\ensuremath{1}}$ for the following combinations,

\begin{eqnarray}
f_{ABC} & = & +1;\nonumber \\
\forall & (ABC) & =(123),\,(471),\,(257),\,(165),\,(624),\,(543),\,(736).\label{eq:44}
\end{eqnarray}
Thus, it is worth noting to suitably handle the octonions in order
to reexamine the $SU(3)_{f}$ symmetry group and its properties.

\section{Generators and Casimir Invariants}

A very useful concept in group representation theory is that of Casimir
operators \cite{key-16,key- 29} although there is no general agreement
\cite{key-15} on a unique definition of Casimir operator\cite{key-8,key-9}.
Nonetheless, there exists a set of invariant matrices \cite{key-15}
which can be constructed from the contractions of the generators of
a Lie group. Such invariant matrices are called the Casimir operators
\cite{key-15,key- 29} and hence commute with all the generators of
a Lie group. A familiar example, from ordinary quantum mechanics,
is the Casimir operator $\hat{L}^{2}$ for the rotation group. It
commutes with the three generators $\overrightarrow{L}$ of the rotation
group and its eigenvalues $l(l+1)\hslash^{2}$ label the 'simplest'
states of particles with angular momentum. Hence, a Casimir operator
is a nonlinear function of the generators that commutes with all of
the generators of the group. The adjoint representation of the algebra
is given by the structure constants , which are always real i.e. 

\begin{align}
\left[\widehat{X}_{A}\,,\widehat{X}_{B}\right]=i & f_{ABC}\,\widehat{X}_{C};\nonumber \\
\left[\widehat{X}_{A}\,,\left[\widehat{X}_{B}\,,\widehat{X}_{C}\right]\right]=if_{BCD}\left[\widehat{X}_{A}\,,\widehat{X}_{D}\right] & =-f_{BCD}f_{ADE}\widehat{X}_{E}.\label{eq:45}
\end{align}
If there are $N$ generators, then we get a $N\times N$ matrix representation
in the adjoint representation. From the Jacobi identity,

\begin{align}
\left[\widehat{X}_{A}\,,\left[\widehat{X}_{B}\,,\widehat{X}_{C}\right]\right]+\left[\widehat{X}_{B}\,,\left[\widehat{X}_{C}\,,\widehat{X}_{A}\right]\right]+\left[\widehat{X}_{C}\,,\left[\widehat{X}_{A}\,,\widehat{X}_{B}\right]\right] & =0;\label{eq:46}
\end{align}
we get the similar relation for the structure constants i.e. $f_{BCD}f_{ADE}+f_{ABD}f_{CDE}+f_{CAD}f_{BDE}=0.$
The states of the adjoint representation correspond to the generators
$|\widehat{X}_{a}>$. We may now consider the simple Lie-algebras
whose generators satisfy the commutation relation

\begin{align}
\left[\mathbb{\mathbb{T}}^{a},\mathbb{\mathbb{T}}^{b}\right]= & \sum_{k}f^{abc}\mathbb{\mathbb{T}}^{c};\label{eq:47}
\end{align}
where we use Hermitian generators $\mathbb{\mathbb{T}}^{a}$ in order
to choose the Killing form \cite{key-16} proportional to $\delta^{ab}$i.e.

\begin{align}
Trace\,\mathbb{\mathbb{T}}^{a}\mathbb{\mathbb{T}}^{b} & \propto\delta^{ab};\label{eq:48}
\end{align}
with a positive and representation dependent proportionality constant
that can be fixed from the choice of the group. With this convention
the structure constants $f^{abc}$ are real and completely anti-symmetric
and the generators of the adjoint representation are related \cite{key-16}
to the structure constants

\begin{align}
\left(\mathbb{\mathbb{T}}_{A}\right)_{bc}^{a}= & -i\, f^{abc}.\label{eq:49}
\end{align}
From the commutation relations (\ref{eq:45}), one can write the anti-symmetric
structure constant as

\begin{align}
f^{abc}= & -\frac{i}{\hat{C}}\,\, Trace\left[\left[\mathbb{\mathbb{T}}^{a},\mathbb{\mathbb{T}}^{b}\right]\mathbb{\mathbb{T}}^{c}\right];\label{eq:50}
\end{align}
where the generators $\mathbb{\mathbb{T}}^{a}$ act on $N\times N$
matrices of a semi simple Lie (Special Unitary $SU(N)$) group of
$N^{2}-1$ generators under the condition $Trace\,\mathbb{\mathbb{T}}^{a}=0$
and $\hat{C}$ is Casimir operator. So, the adjoint representation
of the group plays an important role to examine the Casimir operators
due to the equation (\ref{eq:47}). Hence, the commutation relation
(\ref{eq:46}) does not really diminish the number of generators of
the adjoint representation. Rather, it is a different way of writing
the Jacobi identity \cite{key-16}. Casimir operator for corresponding
group, every expression $\hat{C}$ in the $\mathbb{\mathbb{T}}^{a}$'s
commutes with all the basis elements of the algebra i.e. $\left[\hat{C},\mathbb{\mathbb{T}}^{a}\right]=0.$
In general,$\hat{C}$ does not belong to the algebra as it is not
linear in $\mathbb{\mathbb{T}}^{a}.$ The number of generators required
to give a complete set of Casimir invariants is equal to the rank
of the group. Casimir operators are used to label irreducible representations
of the Lie algebra. The utility of Casimir operators arises from the
fact that all states in a given representation assume the same value
for a Casimir operator. This is because the states in a given representation
are connected by the action of the generators of the Lie algebra and
such generators commute with the Casimir operators. This property
may then be used to label representations in terms of the values of
the Casimir operators.

\section{Casimir Operator for Quaternion $SU(2)$ group}

In Section-2, we have already stated the correspondence between the
quaternion units $e_{j}$ and generators of $SU(2)$ isotopic spin
group. One of the important property of $SU(2)$group is the existence
of invariant operator (namely the total angular momentum, $\hat{\mathbb{J}}^{2}$
in internal isotopic spin space) which commutes with all of the generators
of $SU(2)$ group. For compact groups, the Killing form is just the
Kronecker delta. So, for $SU(2)$ group, the Casimir invariant is
then simply the sum of the square of the generators $\widehat{\mathbb{J}}_{x},$$\widehat{\mathbb{J}}_{y};$$\widehat{\mathbb{J}}_{z}$
of the algebra. i.e., the Casimir invariant is given by 

\begin{align}
\hat{\mathbb{J}}^{2}= & \widehat{\mathbb{J}}_{x}^{2}+\widehat{\mathbb{J}}_{y}^{2}+\widehat{\mathbb{J}}_{z}^{2}.\label{eq:51}
\end{align}
The Casimir eigenvalue in a irreducible representation is $j^{2}=j(j+1)$
where for brevity we use the natural units $c=\hslash=1$. Since the
number of Casimir operators corresponds to the rank of the group,
there is only one Casimir operator in $SU(2)$ which obviously commutes
with all the generators. The Casimir operator is proportional to the
identity element (i.e. the scalar quaternion unit $e_{0}$). This
constant of proportionality can be used to classify the representations
of the Lie group and is also related to the mass or (iso)spin. The
proportional constant for (iso) spin $SU(2)$ group denotes the \textquotedbl{}square\textquotedbl{}
of total (iso)spin $I^{2}=I(I+1)$ for $I=1/2$. We have already stated
that the generators of $SU(2)$ isospin \cite{key-19} in terms of
quaternions are $\hat{T_{j}}=\frac{1}{2}\tau_{j}\equiv\frac{1}{2}\, e_{j}(\forall j=1,2,3)$
and satisfy the commutation relation (\ref{eq:14}) analogous to Eq.
(\ref{eq:47}). So, the Casimir operator for $SU(2)$ quaternion group
is 

\begin{align}
\hat{\mathbb{I}}^{2}= & \widehat{\mathbb{I}}_{x}^{2}+\widehat{\mathbb{I}}_{y}^{2}+\widehat{\mathbb{I}}_{z}^{2}\label{eq:52}
\end{align}
with eigenvalues $I(I+1).$ It is isomorphic to the For $SO(3)$ group
for which the Casimir Operator is the total angular momentum $\hat{\mathbb{J}}^{2}$
given by equation (\ref{eq:51}). So, the members of the irreducible
representations of isospin (or angular momentum) are represented by
the $2$ numbers $I$ and $\mathfrak{m}$ associated respectively
r with the total (iso) spin and its $z-$component i.e.

\begin{align}
\hat{\mathbb{I}}^{2}|I,\,\mathfrak{m}>= & I(I+1)\,|I,\,\mathfrak{m}>\nonumber \\
\widehat{\mathbb{I}_{z}}|I,\,\mathfrak{m}>= & \mathfrak{m}\,|I,\,\mathfrak{m}>.\label{eq:53}
\end{align}
The individual states are connected by the \textquotedblleft{}ladder
operators\textquotedblright{}

\begin{align}
\widehat{\mathbb{I}}_{\pm}= & \widehat{\mathbb{I}_{x}}\pm\widehat{\mathbb{I}_{y}};\nonumber \\
\widehat{\mathbb{I}}_{\pm}|I,\,\mathfrak{m}>= & \sqrt{(I\mp\mathfrak{m})(I\pm\mathfrak{m}+1)}\,|I,\,\mathfrak{m}\pm1>;\label{eq:54}
\end{align}
where $I$ can be determined by the formula $n=2I+1$ ($n$ is the
dimension of the generator) and for a given $I,$ the value of $\mathfrak{m}$
can be $-I,-I+1,...,I-1,I;$ so that it has $2I+1$ degenerate states
for a given $I$. For $SU(2);$ $n=2,\, I=1/2$, and $\mathfrak{m}=-\frac{1}{2}$
or $\mathfrak{m}$=$+\frac{1}{2}$ (an iso doublet). For continuous
symmetries, the resulting quantum numbers (eigenvalues of the \textquotedblleft{}total\textquotedblright{}
Casimir operator \textendash{} the total isospin) are obtained from
the appropriate \textquotedblleft{}addition\textquotedblright{} of
the quantum numbers of the individual representations being added.
The rank of $SU(2)$ group is $n-1=2-1=1$ leading to the definition
of Casimir Operator that there exists only one non linear invariant
operator (i.e. the Casimir Operator) for quaternion $SU(2)$ group.
Hence, the Casimir Operator for $SU(2)$ group may then be constructed
from the square of the isospin (\ref{eq:52}) as 

\begin{align}
\mathsf{\widehat{C}}= & \sum_{a=1}^{3}I_{a}^{2}=-\frac{1}{4}\sum_{a=1}^{3}e_{a}^{2}=-\frac{1}{4}(e_{1}^{2}+e_{2}^{2}+e_{3}^{2})\cong\frac{3}{4}\, e_{0}\equiv\frac{3}{4};\label{eq:55}
\end{align}
where $e_{1}^{2}=e_{2}^{2}=e_{3}^{2}=-1;$ $e_{0}^{2}=e_{0}=1$. is
the unique Casimir operator. This invariant operator obviously commutes
with all the generators of quaternion $SU(2)$ group i.e. 

\begin{align}
\left[\mathsf{\widehat{C}}\,,\widehat{\mathbb{I}_{x}}\right]=\left[\mathsf{\widehat{C}}\,,\widehat{\mathbb{I}_{y}}\right]=\left[\mathsf{\widehat{C}}\,,\widehat{\mathbb{I}_{z}}\right] & =\left[\mathsf{\widehat{C}}\,,\widehat{\mathbb{I}_{\pm}}\right]=0.\label{eq:56}
\end{align}

\section{Split Octonions}

Let us start with an octonion $\mathcal{O}$, which is expanded \cite{key-30}
in a basis $(e_{0},e_{A})$ as 

\begin{align}
\mathcal{O=O}^{0}+\sum_{A=1}^{A=7} & O^{A}e_{A}(\forall A,B.C,=1,2,3,....,7);\label{eq:57}
\end{align}
where $e_{A}(A=1,2,.....,7)$ : the imaginary octonion units and $(e_{0}=1)$
: the real octonion unit satisfy the properties given by Eqs. (\ref{eq:43})
and (\ref{eq:44}) for non-commutative and non-associative octonion
algebra. So, we write the octonion conjugation $(\mathcal{\bar{O}})$,
octonion norm $\left|\mathcal{O}\right|$ and octonion inverse $(\mathcal{O}^{-1})$
\cite{key-6,key- 17,key-18,key-19,key-20} as

\begin{align}
\mathcal{\overline{O}=O}^{0}-\sum_{A=1}^{A=7} & O^{A}e_{A}(\forall A,B.C,=1,2,3,....,7);\label{eq:58}\\
\left|\mathcal{O}\right|=\mathbb{N\,}(\mathcal{O})=<\mathcal{O}|\mathcal{O}>= & \overline{\mathcal{O}}\mathcal{O}=\mathcal{O}\mathcal{\overline{O}}=(\mathcal{O}^{0})^{2}+\sum_{A=1}^{A=7}(\mathcal{O}^{A})^{2};\label{eq:59}\\
\mathcal{O}^{-1}= & \frac{\mathcal{\overline{O}}}{\left|\mathcal{O}\right|}=\frac{\mathcal{\overline{O}}}{\mathbb{N\,}(\mathcal{O})};\label{eq:60}\\
\mathcal{O}^{-1}\mathcal{O}=\mathcal{O}\mathcal{O}^{-1} & =1.\label{eq:61}
\end{align}
For three octonions $x,y,z\in\mathcal{O}$, the non-vanishing associator
is defined by

\begin{align}
<x,\, y,\, z>= & x\,(y\, z)-(x\, y)\, z.\label{eq:62}
\end{align}
It vanishes only for those non-commutating combinations for which
the structure constant $f^{ABC}=1$ i.e. 

\begin{align}
<e_{A},e_{B},\, e_{C}>= & (e_{1},e_{2},e_{3})=(e_{4},e_{7},e_{1})=(e_{2},e_{5},e_{7})=(e_{1},e_{6},e_{5})\nonumber \\
= & (e_{6},e_{2},e_{4})=(e_{5},e_{4},e_{3})=(e_{7},e_{3},e_{6})=0.\label{eq:63}
\end{align}
It justifies that an octonion resembles to $SU(3)$ symmetry consisting
seven non Abelian $SU(2)$ symmetry groups analogous to quaternions.
However, the octonion algebra over the field of complex numbers is
visualized as the Split Octonion algebra \cite{key-1,key- 17} with
its split base units defined as 

\begin{gather}
u_{0}=\frac{1}{2}\left(e_{0}+ie_{7}\right);\,\,\,\,\,\, u_{0}^{\star}=\frac{1}{2}\left(e_{0}-ie_{7}\right);\nonumber \\
u_{m}=\frac{1}{2}\left(e_{m}+ie_{m+3}\right);\,\,\,\,\, u_{m}^{\star}=\frac{1}{2}\left(e_{m}+ie_{m+3}\right)(\forall m=1,2,3);\label{eq:64}
\end{gather}
where $(\star)$ denotes the complex conjugation and $(i=\sqrt{-1})$
(the imaginary unit) commutes with all $e_{A}\,(\forall\, A=1,2,3,....,7)$.
In equation (\ref{eq:64}) $u_{0},\, u_{0}^{\star},\, u_{j},\, u_{j}^{\star}$
are defined as the bi-valued representations of quaternion units $\begin{array}{c}
e_{0}\end{array},\: e_{1},\: e_{2},\: e_{3}$ satisfy $e_{j}e_{k}=-\delta_{jk}+\epsilon_{jkl}e_{l}\,\,(\forall j,k,l=1,2,3).$
Thus, The split octonion basis elements (\ref{eq:64}) satisfy the
following multiplication rules

\begin{align}
u_{i}u_{j}=-u_{j}u_{i}=\epsilon_{ijk}u_{k}^{\star},\,\,\,\,\,\, & u_{i}u_{j}=-u_{j}^{\star}u_{i}^{\star}=\epsilon_{ijk}u_{k};\nonumber \\
u_{i}u_{j}^{\star}=-\delta_{ij}u_{0},\,\,\,\,\,\,\,\,\,\,\,\,\,\,\,\, & u_{i}^{\star}u_{j}=-\delta_{ij}u_{0}^{\star};\nonumber \\
u_{0}u_{i}=u_{i}u_{0}^{\star}=u_{i},\,\,\,\,\,\,\,\,\,\,\,\,\, & u_{0}^{\star}u_{i}^{\star}=u_{i}^{\star}u_{0}=u_{i}^{\star};\nonumber \\
u_{i}u_{0}=u_{0}u_{i}^{\star}=0,\,\,\,\,\,\,\,\,\,\,\,\,\, & u_{i}^{\star}u_{0}^{\star}=u_{0}^{\star}u_{i}=0;\nonumber \\
u_{0}u_{0}^{\star}=u_{0}u_{0}^{\star}=0\,\,\,\,\,\,\,\,\,\,\,\,\, & u_{0}^{2}=u_{0},\,\, u_{0}^{\star2}=u_{0}^{\star}.\label{eq:65}
\end{align}
These relations $\left(\ref{eq:65}\right)$ are invariant \cite{key- 17}
under $G_{2}$ group of automorphism of octonions. Under the $SU(3)$
subgroup of automorphism group $G_{2}$ that leaves the imaginary
unit $e_{7}$( or equivalently the idempotents $u_{0}$and $u_{0}^{\star}$)
invariant, and the $u_{m}$and $u_{m}^{\star}(\forall m=1,2,3)$ transform
like \cite{key- 17} a triplet $(\mathbf{3})$ and anti-triplet $(\mathbf{3^{\star}})$
of $SU(3)_{f}$ . Unlike octonions, the split octonion algebra contains
zero divisors and is therefore not a division algebra.

\section{Split Octonions and $SU(3)_{f}$ Symmetry }

The split octonion algebra may now be regarded as the Lie algebra
of $SU(3)$. So, we may write the complex Octonion $\lyxmathsym{\textgreek{y}}$
in terms of split octonion basis as,

\begin{align}
\psi= & u^{\dagger}\phi+u_{0}^{\star}\phi_{0};\label{eq:66}
\end{align}
where 
\begin{align}
u^{\dagger}= & \left(u_{1}^{\star}u_{2}^{\star}u_{3}^{\star}\right),\,\,\,\phi=\left(\begin{array}{c}
\phi_{1}\\
\phi_{2}\\
\phi_{3}
\end{array}\right).\label{eq:67}
\end{align}
Here, we may illustrate the transformations of the associated split
octonion $u$ with Gell Mann $\lambda$ matrices as,

\begin{align}
u^{\dagger}\lambda_{1}u= & u_{1}^{\star}u_{2}+u_{2}^{\star}u_{1};\nonumber \\
u^{\dagger}\lambda_{2}u= & -i\left(u_{1}^{\star}u_{2}-u_{2}^{\star}u_{1}\right);\nonumber \\
u^{\dagger}\lambda_{3}u= & u_{1}^{\star}u_{1}-u_{2}^{\star}u_{2}=-u_{1}^{\star}\lambda_{8}u=-u^{\dagger}Yu;\nonumber \\
u^{\dagger}\lambda_{4}u= & u_{3}^{\star}u_{1}+u_{1}^{\star}u_{3};\nonumber \\
u^{\dagger}\lambda_{5}u= & -i\left(u_{3}^{\star}u_{1}-u_{1}^{\star}u_{3}\right);\nonumber \\
u^{\dagger}\lambda_{6}u= & u_{2}^{\star}u_{3}+u_{3}^{\star}u_{2};\nonumber \\
u^{\dagger}\lambda_{7}u= & -i\left(u_{2}^{\star}u_{3}-u_{3}^{\star}u_{2}\right);\nonumber \\
u^{\dagger}\lambda_{8}u= & u_{1}^{\star}u_{1}+u_{2}^{\star}u_{2}-2u_{3}^{\star}u_{3}.\label{eq:68}
\end{align}
and
\begin{align}
\frac{1}{\sqrt{3}}u^{\dagger}\lambda_{3}u=\frac{1}{\sqrt{3}}\left(u_{1}^{\star}u_{1}-u_{2}^{\star}u_{2}\right) & =-u_{1}^{\star}\lambda_{8}u=-u^{\dagger}Yu;\nonumber \\
u^{\dagger}\frac{1}{2}\left(\lambda_{3}+\frac{1}{\sqrt{3}}\lambda_{8}\right)u= & u^{\dagger}Qu=2\left(u_{1}^{\star}u_{1}-u_{3}u_{3}^{\star}\right).\label{eq:69}
\end{align}
As such, we may establish the step up ( shift ) operators as

\begin{align}
u^{\dagger}\left(\lambda_{1}+i\lambda_{2}\right)u= & u^{\dagger}I_{+}u=u_{1}^{\star}u_{2};\nonumber \\
u^{\dagger}\left(\lambda_{4}+i\lambda_{5}\right)u= & u^{\dagger}U_{+}u=u_{3}^{\star}u_{1};\nonumber \\
u^{\dagger}\left(\lambda_{6}+i\lambda_{7}\right)u= & u^{\dagger}V_{+}u=u_{2}^{\star}u_{3}.\label{eq:70}
\end{align}
and we may construct the two such independent Casimir (operators)
for the split octonion $SU(3)$ group.Similarly the step down (shift)
operators may be expressed as

\begin{align}
u^{\dagger}\left(\lambda_{1}-i\lambda_{2}\right)u= & u^{\dagger}I_{-}u=u_{2}^{\star}u_{1};\nonumber \\
u^{\dagger}\left(\lambda_{4}-i\lambda_{5}\right)u= & u^{\dagger}U_{-}u=u_{1}^{\star}u_{3};\nonumber \\
u^{\dagger}\left(\lambda_{6}-i\lambda_{7}\right)u= & u^{\dagger}V_{-}u=u_{3}^{\star}u_{2}.\label{eq:71}
\end{align}
The split octonion group thus decomposes $SU(3)$ into singlets ($u_{0}\,;\, u_{0}^{\star})$,
a triplet $u_{i}$ and an anti- triplet $u_{i}^{\star}$. Consequently,
the $SU(3)$ flavor symmetries are suitably handled with split octonions
and may be explored with $I-$, $U-$ and $V-$ spins (flavors) of
$SU(2)$ group as

\begin{eqnarray}
I_{1}=u_{1}^{\star}u_{2}+u_{2}^{\star}u_{1}; & I_{2}=-i\left(u_{1}^{\star}u_{2}-u_{2}^{\star}u_{1}\right); & I_{3}=u_{1}^{\star}u_{1}-u_{2}^{\star}u_{2}\,(I-\, Spin);\label{eq:72}\\
V_{1}=u_{3}^{\star}u_{1}+u_{1}^{\star}u_{3}; & V_{2}=-i\left(u_{3}^{\star}u_{1}-u_{1}^{\star}u_{3}\right); & V_{3}=-i(u_{2}^{\star}u_{2}-u_{3}^{\star}u_{3})\,(V-\, Spin);\label{eq:73}\\
U_{1}=u_{2}^{\star}u_{3}+u_{3}^{\star}u_{2}; & U_{2}=-i\left(u_{2}^{\star}u_{3}-u_{3}^{\star}u_{2}\right); & U_{3}=-i(\left(u_{1}^{\star}u_{1}-2u_{3}^{\star}u_{3}\right))\,(U-\, Spin);\label{eq:74}
\end{eqnarray}
along with the spin octonion valued hyper charge is described as 
\begin{equation}
Y=\frac{1}{\sqrt{3}}\lambda_{8}=\frac{1}{\sqrt{3}}\left(u_{1}^{\star}u_{1}+u_{2}^{\star}u_{2}-2u_{3}^{\star}u_{3}\right).\label{eq:75}
\end{equation}
Likewise, the spin octonion valued shift operators for $I-$, $U-$
and $V-$ spins (flavors) of $SU(2)$ group are also analyzed as
\begin{align}
I_{+}= & u_{1}^{\star}u_{2};\,\,\,\,\,\,\, I_{-}=u_{2}^{\star}u_{1};\nonumber \\
V_{+}= & u_{3}^{\star}u_{1};\,\,\,\,\,\,\, V_{-}=u_{1}^{\star}u_{3}\nonumber \\
U_{+}= & u_{2}^{\star}u_{3};\,\,\,\,\,\,\, U_{-}=u_{2}^{\star}u_{3}.\label{eq:76}
\end{align}
The commutation relations between $I_{+}$ and $I_{-}$,\ $U_{+}$
and $U_{-}$ and $V_{+}$ and $V_{-}$ are then be described

\begin{align}
\left[I_{+},\, I_{-}\right]=2\, I_{3};\,\,\,\,\,\, & \left[I_{3},\, I_{\pm}\right]=\pm I_{\pm};\nonumber \\
\left[U_{+},\, U_{-}\right]=2\, U_{3};\,\,\,\,\, & \left[U_{3},\, U_{\pm}\right]=\pm U_{\pm};\nonumber \\
\left[V_{+},\, V_{-}\right]=2\, V_{3}\,\,\,\,\,\,\, & \left[V_{3},\, V_{\pm}\right]=\pm V_{\pm};\nonumber \\
\left[I_{+},\, V_{+}\right]=\left[I_{+},U_{+}\right]=\left[U_{+},V_{+}\right]= & 0;\nonumber \\
\left[Y,\, I_{3}\right]=\left[Y,\, U_{3}\right]=\left[Y,\, V_{3}\right]= & 0.\label{eq:77}
\end{align}
Accordingly, the Charge $Q$ is described as

\begin{align}
Q= & 2\left(u_{1}^{\star}u_{1}-u_{3}u_{3}^{\star}\right).\label{eq:78}
\end{align}
These equations satisfy all the commutation relations of $I$, $U$
and $V$ - spin multiplets of $SU(3)$ flavor group along with the
Gell- Mann Nishimija relation $Q=\frac{Y}{2}+I_{3}.$

\section{Casimir Operators for split Octonion $SU(3)$ group}

The group $SU(3)$ deals the global $SU(3)$ of quark flavors and
the local $SU(3)_{C}$gauge symmetry of quantum Chromodynamics (QCD).
Here we are interested in the former case where the fundamental representation
is the triplet of (lowest mass) quarks flavors $u$, $d$,$s$. In
the latter case the fundamental representation is the triplet, $A=1,2,3$
(or red, green, blue), with one such triplet for each of the$6$ quark
flavors. Let us start with suitably handled split octonion $SU(3)$
symmetry associated with triplet of (lowest mass) quarks flavors $u$,
$d$,$s$. So, formulas (\ref{eq:53}) and (\ref{eq:54}) can be generalized
to the case of $SU(3)$ symmetry where we have $n=3$, $I=1$, and
$\mathfrak{m}=+1,0,-1$ forming an iso-triplet of triplet of quarks
flavors $u$, $d$,$s$. The group $SU(3)$ is of rank two (i.e. $n-1=3-1=2)$.
So, it is to be noted that the individual members of an irreducible
representation of $SU(3)$ are labeled by 2 constants. These are defined
by the third component of Iso-spin $\mathbb{\widehat{I}}_{3}=\frac{1}{2}\lambda_{3}$
and hypercharge $\widehat{Y}=\frac{1}{2\sqrt{3}}\lambda_{8}$both
are diagonal. Consequently, we may construct \cite{key-13} two independent
the Casimir invariant tensors (operators) for split Octonion $SU(3)$
group as 
\begin{align}
\boldsymbol{\widehat{C}}_{2}=-\frac{2}{3}\, i\, f_{ABC}\,\widehat{F_{A}}\widehat{\, F_{B}}\,\widehat{F_{C}}= & \sum_{A=1}^{A=8}\widehat{F}_{A}^{2}\label{eq:79}
\end{align}
and
\begin{align}
\boldsymbol{\widehat{C}}_{3}= & d_{ABC}\widehat{F_{A}}\widehat{\, F_{B}}\,\widehat{F_{C}}\label{eq:80}
\end{align}
where $F_{A}\,(\forall A,B,C=1,2,3.....,8)$ are suitably handled
with split octonions in terms of $I-$, $U-$, and $V-$ spins $SU(3)$
flavor symmetry. It is easily verified that $\boldsymbol{\widehat{C}}_{2}$
and $\boldsymbol{\widehat{C}}_{3}$ commute with all the generators
$\widehat{F}_{A}$ and the shift operators $I-$, $U-$, and $V-$
spins of split octonion $SU(3)$ group as well. These Casimir operators
$\boldsymbol{\widehat{C}}_{p}(\forall p=2,3)$ also commute with the
Hamiltonian $\hat{H}$ of strong interactions.

\section{Conclusion}

The foregoing analysis provides a fundamental representation of quaternions
and the split octonion algebra in order to investigate the flavor
$SU(2)$ and $SU(3)$ symmetries. Instead of Pauli Spin matrices for
$SU(2)$ group we have used the compact notations of quaternions while
for the case of $SU(3)$ symmetry we have established the connection
between the Gellmann $\lambda$ matrices and octonion basis elements.
The isospin symmetry and the quark-anti quark symmetries are suitably
handled with quaternions and octonions. It is shown that the structure
constants of $SU(3)$ symmetry resembles with those of multiplication
identities of octonion basis elements. This result is useful for the
use of familiar table of (Clebsch-Gordan) angular momentum coupling
coefficients for combining quark and anti-quark states together by
using quaternions and octonions. It is remarkable that the $SU(3)$
group of split octonion is an invariant sub group of octonion automorphism
group $G_{2}$. So, it is necessary to handle whole $SU(3)$ symmetry
in terms of split octonions in an enthusiastic manner. Since Casimir
operators for split octonion gives explicit basis for various sub-algebras,
a set of commuting operators in terms of various permutations of octonions
and isospin multiplets is the useful for the unique and consistent
representation for $SU(2)$ and $SU(3)$ flavor symmetries and the
theory of strong interactions as well. Consequently, the analogous
Casimir operators for $SU(2)$ and $SU(3)$ flour symmetry groups
are analyzed and suitably handled respectively with quaternions and
octonions. It is also shown that analogous Casimir operators commute
with the corresponding generators of $SU(2)$ and $SU(3)$ gauge groups.

\textbf{ACKNOWLEDGMENT}: One of us (OPSN) acknowledges the financial
support from Third World Academy of Sciences, Trieste (Italy) and
Chinese Academy of Sciences, Beijing under UNESCO-TWAS Associateship
Scheme. He is also thankful to Prof. Yue-Liang Wu for his hospitality
and research facilities\textbf{ }at ITP and KITPC, Beijing (China).

\end{document}